\documentclass[prd,twocolumn,showpacs,superscriptaddress,nofootinbib,floatfix,showkeys,10pt]{revtex4-1}
\usepackage{graphicx}
\usepackage{amsmath}
\usepackage{bm}
\usepackage{yhmath}
\usepackage{mathtools}
\usepackage{wasysym}
\usepackage[colorlinks,citecolor=blue,urlcolor=blue,linkcolor=blue]{hyperref}
\usepackage{subfigure}
\usepackage{color}
\usepackage{cases}
\usepackage{subfigure}
\usepackage{times}
\usepackage{dcolumn,booktabs,bm}
\usepackage{slashed}
\usepackage{amsfonts,amssymb,stmaryrd,latexsym,amsmath}
\usepackage{textcomp}
\usepackage{multirow}
\usepackage{cancel}
\usepackage{array}

\def\SU3{{\text{SU(3)}_{\rm F}}}

\allowdisplaybreaks

\begin{document}

\title{ Probing the long-range structure of the $T_{cc}^+$ with the strong and electromagnetic decays}

\author{Lu Meng}
\affiliation{Ruhr-Universit\"at Bochum, Fakult\"at f\"ur Physik und
Astronomie, Institut f\"ur Theoretische Physik II, D-44780 Bochum,
Germany }

\author{Guang-Juan Wang}
\affiliation{Advanced Science Research Center, Japan Atomic Energy
Agency, Tokai, Ibaraki, 319-1195, Japan}

\author{Bo Wang}\email{wangbo@hbu.edu.cn, corresponding author}
\affiliation{School of Physical Science and Technology, Hebei
    University, Baoding 071002, China}
\affiliation{Key Laboratory of High-precision Computation and
Application of Quantum Field Theory of Hebei Province, Baoding
071002, China}

\author{Shi-Lin Zhu}\email{zhusl@pku.edu.cn, corresponding author}
\affiliation{School of Physics and Center of High Energy Physics,
Peking University, Beijing 100871, China}

\begin{abstract}
Very recently, the LHCb Collaboration reported the doubly charmed
tetraquark state $T_{cc}^+$ below the  $D^{*+}D^0$ threshold about
$273$ keV. As a very near-threshold state, its long-distance
structure is very important. In the molecular scheme,  we relate the
coupling constants of $T_{cc}^+$ with $D^{*0}D^+$ and $D^{*+}D^0$ to
its binding energy and mixing angle of two components with a
coupled-channel effective field theory.  With the coupling
constants, we investigate the kinetically allowed strong decays
$T_{cc}^+\to D^0D^0\pi^+$, $T_{cc}^+\to D^+D^0\pi^0$ and radiative
decays $D^+D^0 \gamma$. Our results show that the decay width of
$T_{cc}^+\to D^0D^0\pi^+$ is the largest one, which is just the
experimental observation channel. Our theoretical total strong and
radiative widths are in favor of the $T_{cc}^+$ as a
$|D^{*+}D^0\rangle$ dominated bound state. The total strong and radiative width in the single channel
limit and isospin singlet limit are given as $59.7^{+4.6}_{-4.4} \text{ keV}$ and $46.7^{+2.7}_{-2.9} \text{ keV}$, respectively. Our calculation is cutoff-independent and without
prior isospin assignment. The absolute partial widths and ratios of
the different decay channels can be used to test the structure of
$T_{cc}^+$ state when the updated experimental results are
available.

\end{abstract}

\maketitle

\thispagestyle{empty}

\section{Introduction}

Very recently, the LHCb Collaboration reported the first doubly
charmed tetraquark state $T_{cc}^+$ in the prompt production of the
$pp$ collision with a  signal significance over 10
$\sigma$~\cite{LHCb:2021vvq}. Its mass with respect to the
$D^{*+}D^0$ threshold and width are
\begin{eqnarray}
    \delta m    &=&-273\pm61\pm5_{-14}^{+11}\text{ keV},\nonumber\\
    \Gamma  &=&410\pm165\pm43_{-38}^{+18}\text{ keV}.\label{eq:lhcb_1}
\end{eqnarray}
In the fitting, the quantum number $J^P=1^+$ is assumed.  The
significance for $\delta m<0$ is 4.3 $\sigma$. The LHCb Collaboration also released a decay analysis, in which the unitarised Breit-Wigner profile was used~\cite{LHCb:2021auc}~\footnote{To some extent, our results agree with analysis in Ref.~\cite{LHCb:2021auc}. We should stress that the analysis was released after our work. Our calculation only based on the information in Refs.~\cite{talkeeps1,epstalk2} and was independent on the Ref.~\cite{LHCb:2021auc}. }. The mass  with respect to the
$D^{*+}D^0$ threshold  and width read,
\begin{eqnarray}
\delta m^U &=&-361\pm40 \text{ keV}, \quad \Gamma^U=47.8\pm 1.9 \text{ keV}. \label{eq:lhcb_2}
\end{eqnarray}
 The observation of
$T_{cc}^+$ is a great breakthrough for the hadron physics. It is the
second doubly charmed hadron that has been observed in experiments
for now. What is more interesting, it is manifestly an exotic hadron
composed of four (anti)quarks.

In fact, the doubly heavy tetraquark states are anticipated and
debated for 40
years~\cite{Carlson:1987hh,Silvestre-Brac:1993zem,Semay:1994ht,Pepin:1996id,Gelman:2002wf,Vijande:2003ki,Janc:2004qn,Cui:2006mp,Navarra:2007yw,Vijande:2007rf,Ebert:2007rn,Lee:2009rt,Yang:2009zzp,Du:2012wp,Feng:2013kea,Ikeda:2013vwa}.
In 2017, the first doubly charmed baryon $\Xi_{cc}^{++}$ was
observed by the LHCb Collaboration~\cite{LHCb:2017iph}, which
incited a new round of heated discussions on the doubly heavy
tetraquark
states~\cite{Luo:2017eub,Karliner:2017qjm,Eichten:2017ffp,Wang:2017uld,Cheung:2017tnt,Park:2018wjk,Francis:2018jyb,Junnarkar:2018twb,Deng:2018kly,Yang:2019itm,Liu:2019stu,Tan:2020ldi,Lu:2020rog,Braaten:2020nwp,Gao:2020ogo,Cheng:2020wxa,Noh:2021lqs,Faustov:2021hjs}.
An extensive review of the $T_{cc}$ system can be found in
Ref.~\cite{Liu:2019zoy}. From the theoretical perspective, a
well-known  fascinating  feature of the compact doubly heavy
tetraquark states is that they might locate below the two-meson
thresholds and then become very narrow. The underlying reason is the
possible heavy-antiquark-heavy-diquark symmetry. The doubly heavy
diquark  in color anti-triplet could be relatively compact, and it
is an analog of the antiquark. The mass of doubly heavy compact
tetraquark is constrained by its singly heavy  partner in the
heavy-antiquark-heavy-diquark symmetry (e.g.
see~\cite{Cohen:2006jg,Karliner:2017qjm,Eichten:2017ffp} for
details). The above analyses were well accepted  for doubly bottom
systems due to large bottom quark mass. However, there was no
agreement for doubly charmed systems before the observation of
$T_{cc}^+$ state. 

Apart from the compact tetraquark scheme, there is
another motivation to investigate the doubly heavy tetraquark states
in the hadronic molecule scheme, which might not be as popular as
the former one but has the equal significance. In molecular scheme,
the one-pion-exchange interaction of $\bar{D}^*D/\bar{D}D^*$ system
with the quantum numbers of $I(J^{PC})=0(1^{++})$ [corresponding to
$X(3872)$] and that of the $D^*D$ system with $I(J^P)=0(1^+)$ are
exactly the same in the isospin symmetry
limit~\cite{Dias:2011mi,Li:2012cs,Li:2012ss}. The doubly charmed
analog of $X(3872)$ is therefore
expected~\cite{Dias:2011mi,Li:2012cs,Li:2012ss}. In Ref.~\cite{Li:2012ss}, the authors obtained a $D^*D$ bound state with quantum numbers $I(J^P)=0(1^+)$, in which the long-range one-pion-exchange as well as the short- and mid-range interactions by exchanging $\eta$, $\rho$, $\omega$ and $\sigma$ mesons were included. The theoretical binding energy and root-mean-square radius are 470 keV and 4.46 fm, respectively. The predictions using the one-boson-exchange model agree very well with the
newly experimental results~\cite{LHCb:2021vvq}. Similar results were also obtained in chiral effective field theory~\cite{Xu:2017tsr}. After the observation of $T_{cc}^+$, the isospin violating effect was considered in the one-boson-exchange model~\cite{Chen:2021vhg}. The $T_{cc}^+$ was interpreted as a bound state composed of two channels, $\cos \theta |D^{*+}D^0\rangle +\sin \theta |{D}^{*0}D^+\rangle$ with $\theta \approx \pm 30.08^\circ$.  In this work, we will see
the newly observed $T_{cc}^+$ tetraquark state does have many
similarities with the $X(3872)$.

The $T_{cc}^+$ state is only about 300 keV below the $D^{*+}D^0$
threshold. If the $T_{cc}^+$ is interpreted as the bound state of
$D^{*+}D^0$ in a single channel formalism, a natural consequence of
such a small binding energy is the low-energy universality similar
to the $X(3872)$ state~\cite{Braaten:2003he,Braaten:2004rn}. The
low-energy observables for $T_{cc}^+$ or $X(3872)$ are insensitive
to the details of the interactions. Thus, the long-range feature of
such systems only depends on the scattering length or binding
energy.  If the higher $D^{*0}D^+$ channel was taken into
consideration in a coupled-channel formalism, the considerable
isospin violation effect is expected due to the sensitivity of the
structure to the threshold differences for such a very
near-threshold bound state.  In this work, we aim to uncover the
structure of the $T_{cc}^+$ through its long-distance dynamics, the
strong and radiative decays. We will resort to an effective field
theory satisfying the renormalization group invariance. The coupling
constants of $T_{cc}^+$ with $D^{*+}D^0$ and $D^{*0}D^+$ will be
related to the binding energy and mixing angle of the two
components. The strong and radiative decay widths can provide
important information about its structure.

This work is organized as follows. In section~\ref{sec:cp}, we use a
coupled-channel effective field theory to relate the coupling
constants to the binding energy and mixing angle of the two
components. In section~\ref{sec:strong}, we calculate the strong and
radiation decays for the $T_{cc}^+$ states and provide some insights
into its structure. In section~\ref{sec:sum}, we give a brief
summary.

\section{Coupling constants and wave functions}~\label{sec:cp}
In the molecular scheme, the two closest thresholds are $D^{*+}D^0$
and $D^{*0}D^+$, which are located above the $T_{cc}^+$ about $0.3$
MeV and $1.7$ MeV, respectively. The components of the bound state
will be sensitive to the threshold mass gaps and the large isospin
violation effect is expected~\cite{Li:2012cs}. Therefore, we will
introduce the coupled-channel effect dynamically rather than
presuming a prior isospin assignment.  The two related channels are
noted as  $|1\rangle\equiv| D^{*+}D^0\rangle$ and $|2\rangle
\equiv|D^{*0}D^+\rangle$. We adopt a coupled-channel effective field
theory proposed by Cohen {\it et al.}~\cite{Cohen:2004kf}, which is
well used in  hadron
physics~\cite{Braaten:2005ai,Meng:2020cbk,Dong:2020hxe} and nuclear
physics~\cite{Higa:2020kfs}. We will see that in this effective
field theory, the cutoff-dependence can be eliminated exactly, which
makes it renormalization group invariant.

For the effective field theory, we introduce the leading order
interaction
\begin{equation}
V(\bm{p},\bm{p}')=\left[\begin{array}{cc}
    v_{11} & v_{12}\\
    v_{12} & v_{22}
\end{array}\right]\Theta(\Lambda-p)\Theta(\Lambda-p'),
\end{equation}
where $v_{ij}$ are energy-independent parameters. The step function
$\Theta$  serves as a hard regulator and $\Lambda$ is the cutoff
parameter. For such a separable interaction, the $T(\bm{p},\bm{p'})$
has the similar separable form with
$T(\bm{p},\bm{p'})=t\Theta(\Lambda-p)\Theta(\Lambda-p')$, where $t$
is the matrix of elements $t_{ij}$. The coupled-channel
Lippmann-Schwinger equations (LSEs) can be reduced to a set of
algebraic equations,
\begin{equation}
    t=v+vGt\Longrightarrow t=(1-vG)^{-1}v,
\end{equation}
where $G=\text{diag}\{G_1,G_2\}$. The $G_i$ reads
\begin{equation}
    G_{i}(E)=\int^{\Lambda}\frac{d^{3}\bm{q}}{(2\pi)^{3}}\frac{1}{E-E_{i,q}+i\epsilon},\, E_{i,q}=\delta_{i}+\frac{q^{2}}{2\mu},
\end{equation}
where $\mu$ is the reduced mass of di-mesons. In this work, we can
neglect the tiny differences of the reduced masses in the  two
channels. The $\delta_i$ is the mass difference with respect to
$D^{*+}D^0$ threshold. Then we have $\delta_1=0$   and
$\delta=\delta_2\equiv m_{D^{*0}}+m_{D^+}-(m_{D^{*+}}+m_{D^0})$. For
a bound state with $E<0$ , the explicit expression of $G_i(E)$ reads
\begin{eqnarray}
G_{i}(E)&=\frac{\mu}{\pi^{2}}\left[-\Lambda+k_{i}\arctan\frac{\Lambda}{k_{i}}\right]\approx\frac{\mu}{2\pi}\left[-\frac{2}{\pi}\Lambda+k_{i}\right],\label{eq:G1}
\end{eqnarray}
where $k_{1}\equiv\sqrt{-2\mu E}$ and
$k_{2}\equiv\sqrt{-2\mu(E-\delta)}$. The approximation in
Eq.~\eqref{eq:G1} is a consequence of $|k_i|\ll \Lambda$. It is
straightforward to obtain the solution of the LSEs,
\begin{equation}
t=\frac{1}{D}\left[\begin{array}{cc}
    b_{11}b_{12}^{2}\left(1-b_{22}k_{2}\right) & b_{11}b_{12}b_{22}\\
    b_{11}b_{12}b_{22} & b_{12}^{2}b_{22}\left(1-b_{11}k_{1}\right)
\end{array}\right],\label{eq:tmrx}
\end{equation}
where
$D=\frac{\mu}{2\pi}\left[b_{12}^{2}\left(b_{11}k_{1}-1\right)\left(b_{22}k_{2}-1\right)-b_{11}b_{22}\right]$
and the $b_{ij}$ are introduced as
\begin{eqnarray}
    \begin{cases}
        \frac{1}{b_{11}} & =\frac{2\pi}{\mu}(\frac{v_{22}}{v_{11}v_{22}-v_{12}^{2}}-G_{1})+k_{1}\\
        \frac{1}{b_{22}} & =\frac{2\pi}{\mu}(\frac{v_{11}}{v_{11}v_{22}-v_{12}^{2}}-G_{2})+k_{2}\\
        \frac{1}{b_{12}} & =\frac{2\pi}{\mu}\frac{v_{12}}{v_{11}v_{22}-v_{12}^{2}}
    \end{cases}.~\label{eq:defineb}
\end{eqnarray}
One can see that the cutoff dependence of $G_i$ in
Eq.~\eqref{eq:defineb} can be absorbed by renormalizing $v_{ij}$. In
this way, the cutoff dependence can be eliminated exactly. The
similar results were derived in
Refs.~\cite{Cohen:2004kf,Braaten:2005ai}.

The bound state $T_{cc}^+$ corresponds to a pole in the real axial
of the complex energy plane. The residuals of the $t$ matrix can be
related to the coupling constants of the bound state with the
corresponding di-meson
channels~\cite{Gamermann:2009uq,Gamermann:2009fv}. In our
normalization convention, we have
\begin{equation}
\lim_{E\to E_{0}}(E-E_{0})t_{ij}=\lim_{E\to
E_{0}}\left[\frac{d(t_{ij})^{-1}}{dE}\right]^{-1}=\frac{1}{8M_{T}^{2}\mu}g_{i}g_{j}
\end{equation}
where $E_0$ is the pole corresponding to  the $T_{cc}^+$ state.
$M_T$ is the mass of $T_{cc}^+$ and $g_i$ is its coupling constant
to the two $\bar D D^*$ channels. A straightforward derivation gives
a very simple expression for $\lim_{E\to E_{0}}(E-E_{0})t$, which
reads
\begin{eqnarray}
\frac{2\pi}{\mu^{2}}\left[\begin{array}{cc}
        \kappa_{1}\cos^{2}\theta & \sqrt{\kappa_{1}\kappa_{2}}\sin\theta\cos\theta\\
        \sqrt{\kappa_{1}\kappa_{2}}\sin\theta\cos\theta & \kappa_{2}\sin^{2}\theta
    \end{array}\right],\label{eq:tmx_rsd}
\end{eqnarray}
where $\kappa_i$ and $\theta$ are defined as
\begin{equation}
    \kappa_{i}\equiv\lim_{E\to E_{0}}k_{i},\quad\tan^{2}\theta\equiv\frac{b_{22}\kappa_{1}\left(b_{11}\kappa_{1}-1\right)}{b_{11}\kappa_{2}\left(b_{22}\kappa_{2}-1\right)}.\label{eq:theta}
\end{equation}
We will see that $\theta$ is actually the mixing angle of the two
channels in the $T^+_{cc}$ state. The coupling constants read
\begin{equation}
    g_{1}=\frac{4M_{T}\sqrt{\pi\kappa_{1}}}{\sqrt{\mu}}\cos\theta,\quad g_{2}=\frac{4M_{T}\sqrt{\pi\kappa_{2}}}{\sqrt{\mu}}\sin\theta.~\label{eq:g1g2}
\end{equation}
We want to emphasize  that our results are more general than those
in Refs.~\cite{Gamermann:2009uq,Gamermann:2009fv}, in which
$v_{11}=v_{12}=v_{22}=v$ is assumed.

In principle, the coupling constants obtained from the residuals of
the $T$-matrix can be related to the wave functions for the bound
states~\cite{Braaten:2005ai,Gamermann:2009uq,Gamermann:2009uq,Aceti:2012dd,Sekihara:2016xnq}.
In our interaction the Schr\"odinger equation reads,
\begin{equation}
    (\hat{H}_{0}+\hat{V})|\psi\rangle=E_{0}|\psi\rangle,\quad\hat{V}=\sum_{i,j}\frac{1}{(2\pi)^{3}}v_{ij}|i\rangle\langle j|.
\end{equation}
The solution of the coupled-channel equation can be obtained by the
combination of the single-channel wave functions,
\begin{eqnarray}\label{eq:wvfunc}
    &\langle\bm{p}|\psi\rangle=c_{1}\phi_{1}(p)|1\rangle+c_{2}\phi_{2}(p)|2\rangle,\\
    &\phi_{i}(p)=\xi_{i}\frac{\Theta(\Lambda-p)}{E_{0}-\frac{p^{2}}{2\mu}-\delta_{i}},\quad\xi_{i}^{2}\approx\frac{\kappa_{i}}{4\pi^{2}\mu^{2}},
\end{eqnarray}
where $\xi_i$ is the normalization constant. $c_i$ is the
coefficient of two components and satisfies $c_1^2+c_2^2=1$. For the
$T$-matrix, one can take the approximation near the bound state
pole~\cite{Sekihara:2016xnq},
\begin{equation}
    T_{ij}(\bm{p},\bm{p'})\approx(2\pi)^{3}\frac{\langle\bm{p},i|\hat{V}|\psi\rangle\langle\psi|\hat{V}|\bm{p'},j\rangle}{E-E_{0}},
\end{equation}
where   $\langle\bm{p},i|\hat{V}|\psi\rangle$ can be substituted by
\begin{eqnarray}
    \langle\bm{p},i|\hat{V}|\psi\rangle&=&\langle p,i|\hat{H}-\hat{H}_{0}|\psi\rangle=\left(E_{0}-\frac{p^{2}}{2\mu}-\delta_{i}\right)\langle\bm{p},i|\psi\rangle \nonumber\\
    &=&c_{i}\xi_{i}\Theta(\Lambda-p).
\end{eqnarray}
Therefore, we obtain the element of $T$-matrix,
\begin{equation}
    t_{ij}\approx (2\pi)^{3}\frac{c_{i}c_{j}\xi_{i}\xi_{j}}{E-E_{0}}~\label{eq:tmx_wv}.
\end{equation}
Comparing the above expression with Eq.~\eqref{eq:tmx_rsd}, one can
obtain the meaning of $c_i$,
\begin{equation}
    c_1=\cos \theta,\quad c_2=\sin\theta.
\end{equation}
Thus, we proved that the $\theta$ defined in Eq.~\eqref{eq:theta} is
in fact the mixing angle of the $|1\rangle$ and $|2\rangle$
components.

One can see the coupling constants in Eq.~\eqref{eq:g1g2} depend on
the binding energy (in $\kappa_i$) and the mixing angle $\theta$. In
the single channel limit ($\theta=0$), the coupling constant and the
wave function only depend on the binding energy, which is the
manifestation of the universality of the low energy dynamics. In the
realistic case, the long-range dynamics of $T_{cc}^+$ will rely on
the mixing angle. One can extract the structure information of
$T_{cc}^+$ by investigating its strong and radiative decays.

\section{Strong decay and radiative decay}~\label{sec:strong}

\begin{figure*}[htp]
    \centering  \includegraphics[width=1.0\textwidth]{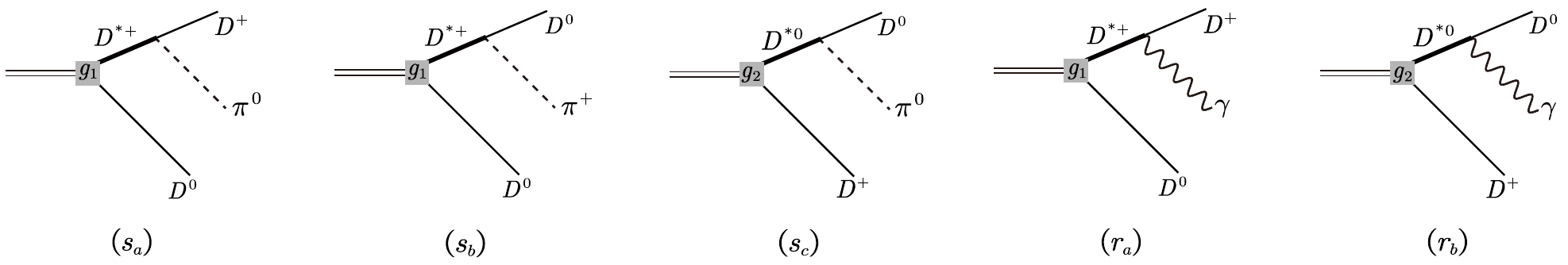}
    \caption{The Feynman diagrams for strong and radiative decays of the $T_{cc}^+$ state, where the vertices marked by $g_1$ and $g_2$ denote the coupling strengths of $T_{cc}^+$ to the channels $|1\rangle$ and $|2\rangle$, respectively.}\label{fig:feyn}
\end{figure*}

The strong and radiative decays of $T_{cc}^+$ state are illustrated
in Fig.~\ref{fig:feyn}. The details for the determinations of the
coupling constants of $D^\ast\to D\pi(\gamma)$, the strong and
radiative decay amplitudes of $T_{cc}^+$, as well as the each
diagram contribution in ideal single-channel cases are given in
Appendix~\ref{app:amplt}. Here, we list some main conclusions that
one can read from Appendix~\ref{app:amplt}. 
The results show that the Figs.~\ref{fig:feyn}($s_b$) and~($r_b$)
are the dominant diagrams contributing to the strong and radiative
decays, respectively, which are almost 4 times larger than the
contributions from other diagrams. For the strong decays, we use ($s_b$) to represent two diagrams considering the exchange of two identical $D^0$ final state. The
amplitude of diagram ($s_b$) is amplified by an extra isospin factor
$\sqrt{2}$ in the $D^{*+} D^0\pi^+$ vertex and considerable interference effect of two diagrams. For
the radiative decays, the amplitude of ($r_b$) is much larger than
that of ($r_a$), because the leading amplitudes for M1 radiative
transition $D^{*0,+}\to D^{0,+}\gamma$ are roughly proportional to
the electric charges of the light quarks in the heavy quark limit.
In addition, the strong decay width arising from ($s_b$) is also
much larger than the radiative one from ($r_b$).

However, the realistic situation is the strong and radiative decay
widths depend on the binding energy of $T_{cc}^+$ as well as the
mixing angle of two components [see Eq.~\eqref{eq:g1g2}]. We list
three special angles and their corresponding states as follows,
\begin{eqnarray}
&|T_{cc}^{+}\rangle=\cos\theta|D^{*+}D^{0},\phi_{1}\rangle+\sin\theta|D^{*0}D^{+},\phi_{2}\rangle,\\
&\theta=\begin{cases}
    0 & \text{pure }D^{*+}D^{0}\\
    \frac{\pi}{4} & I=1,I_{3}=0\\
    -\frac{\pi}{4} & I=0,I_{3}=0
\end{cases}.
\end{eqnarray}

\begin{figure*}[htp]
    \centering  \includegraphics[width=0.33\textwidth]{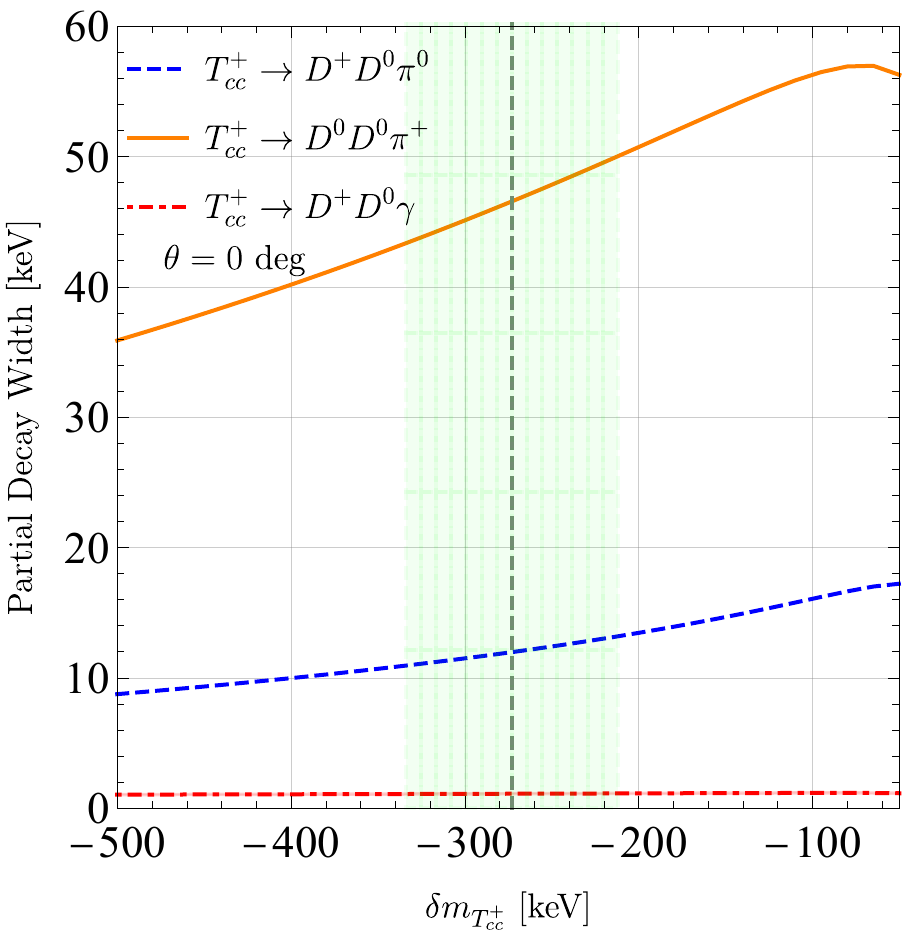}
    \includegraphics[width=0.32\textwidth]{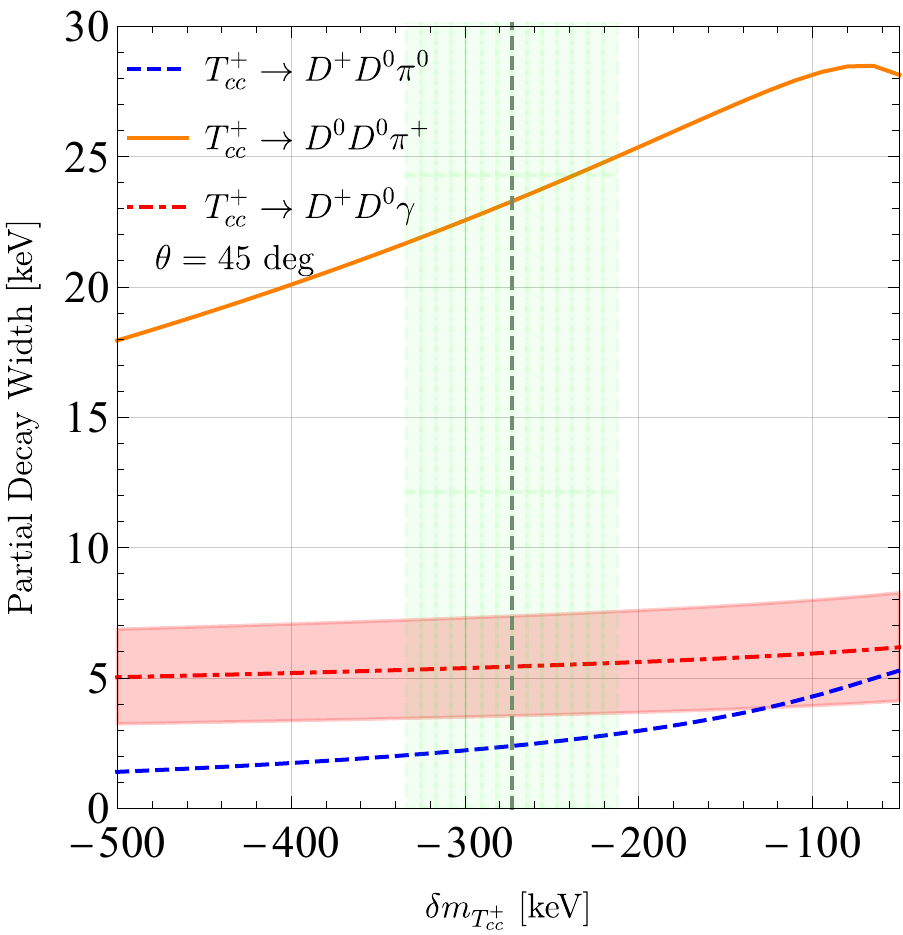}
    \includegraphics[width=0.32\textwidth]{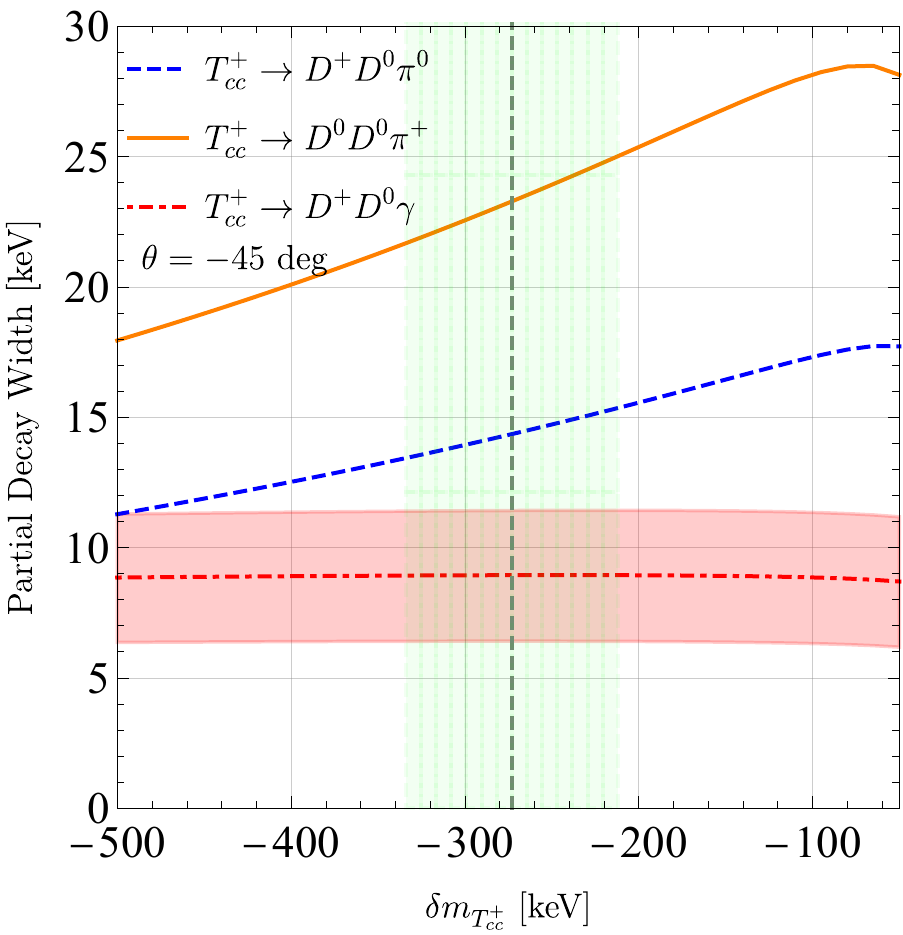}\\
    \caption{The dependence of the strong and radiative decay widths on binding energy for the $T_{cc}^+$ state. Three subfigures from the left to right show  the decay widths of single channel ($D^{*+}D^0$), $I=1$ (isospin triplet) and $I=0$ (isospin singlet) configurations, respectively. The green band represents the uncertainty of the $T_{cc}^+$ mass. The vertical dashed line is the central value of the binding energy. For the radiative decay, the red shadow represents the uncertainties arising from the unfixed $D^{*0}$ width ($40$$-$$80$ keV is used).  }\label{fig:1to3}
\end{figure*}

\begin{figure*}[!htbp]
    \includegraphics[width=0.33\textwidth]{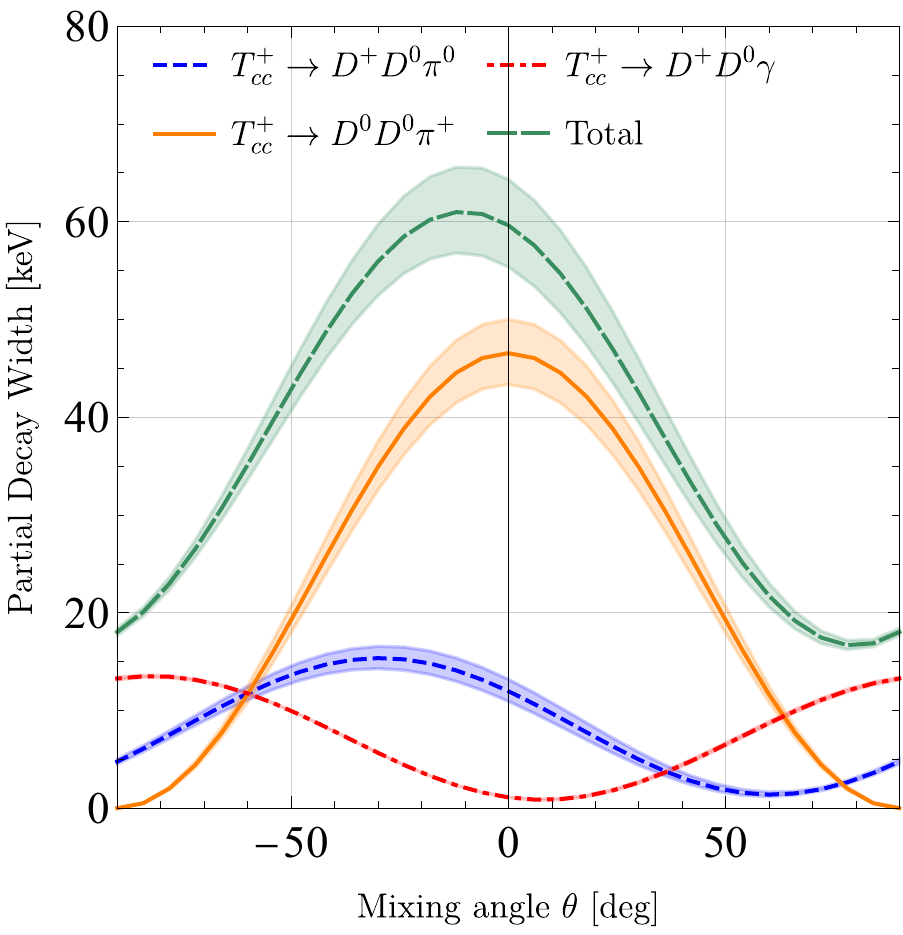}\hspace{2cm}
    \includegraphics[width=0.32\textwidth]{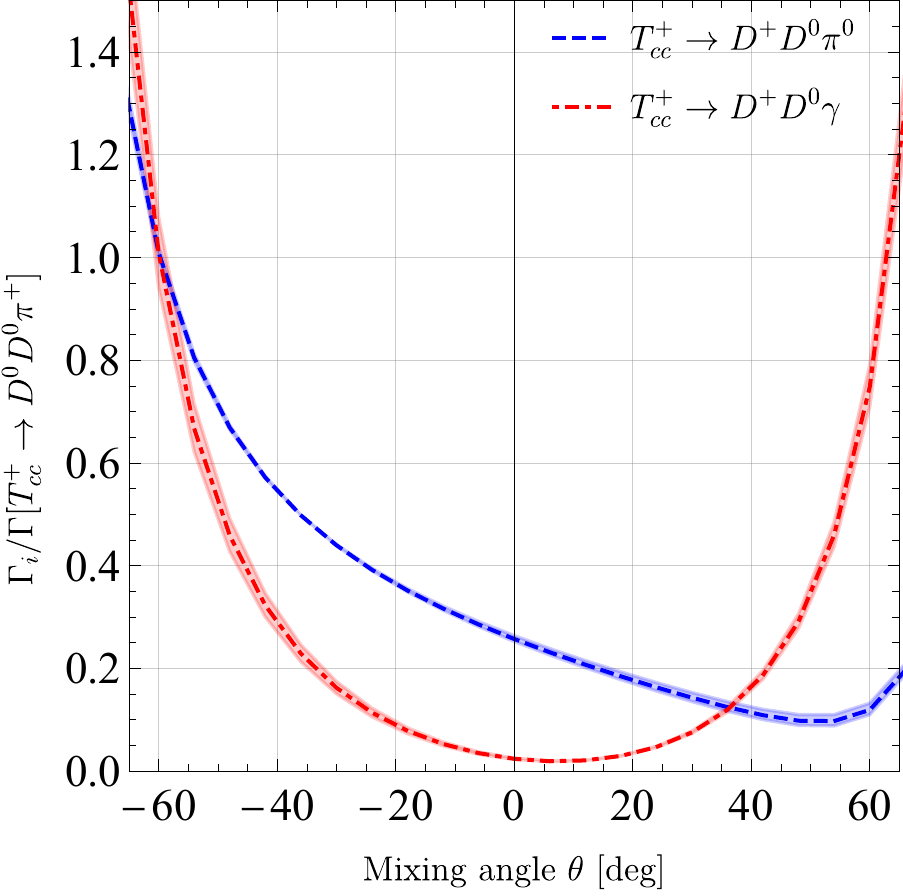}
    \caption{The dependence on the mixing angle for the $T_{cc}^+$ strong and radiative decay. The left and the right subfigure show the absolute values and relative ratios of the decay widths. The colored shadow represents the uncertainties stemming from the $T_{cc}^+$ mass in Eq.~\eqref{eq:lhcb_1}.}\label{fig:4to6}
\end{figure*}

 Either the absolute value or the relative ratios of the partial decay widths,  embed the important information about the structure of the $T_{cc}^+$ state. We present the strong and radiative decay widths in Figs.~\ref{fig:1to3} and ~\ref{fig:4to6}. We can obtain several nontrivial conclusions from them.

The first and the foremost conclusion is that the dominant decay
mode of the $T_{cc}^+$ is $D^0D^0\pi^+$, which is just its
observation channel in experiments. In Fig.~\ref{fig:1to3},  we
present the partial decay widths  with the mixing angle $\theta=0$,
$\pi/4$ and $-\pi/4$, which correspond to the single channel
$D^{*+}D^0$, $I=1$ and $I=0$ cases, respectively. In these three
configurations, the dominant decay mode is $T_{cc}^+\to
D^0D^0\pi^+$. In the left subfigure of Fig.~\ref{fig:4to6}, we show
the dependence of the decay widths on the mixing angle. One can see
that the  $T_{cc}^+\to D^0D^0\pi^+$ is dominant in  most mixing
structures. The exception only appears when the bound state is
almost pure $D^{*0}D^+$ bound state ($|\theta|\sim \pi/2$). But it
is less likely that a bound state (blow two thresholds) in
two-channel interaction model contains more higher channel
component. Thus, in the molecular scheme, it is easy to understand
why the $T_{cc}^+$ is firstly observed in the $D^0D^0\pi^+$ final
state rather than other channels.

Meanwhile, the experimental decay width of $T_{cc}^+$ is in favor of
the $|D^{*+}D^0\rangle$ dominant molecule structure. In the left
subfigure of Fig.~\ref{fig:4to6}, the maximum of the total decay
width appears at $\theta\approx 0$, because the dominant decay mode
$T_{cc}^+\to D^0D^0\pi^+$ is induced by the  $D^{*+}D^0$ channel
through the coupling constant $g_1$ proportional  to
$\text{cos}\theta$. From Eq.~\eqref{eq:G1}, we can see the partial
decay width achieves its maximum when $\theta\approx 0$, which corresponds
to the single channel limit. In this limit, we obtain the total
width of $T_{cc}^+$ from the strong and the radiative decays as
\begin{eqnarray}
\text{Single-channel limit: }\Gamma_{\text{str}}+\Gamma_\text{EM}
=59.7^{+4.6}_{-4.4} \text{ keV}.~\label{eq:wd_single}
\end{eqnarray}
This decay width is still smaller that the central value 410 keV in
experiment. The
difference might be resolved when the experimental resolution is
improved in the future.  The parameters of near-threshold resonance would be sensitive to the lineshape parameterization formalism~\footnote{The analysis from LHCb Collaboration after this work with unitarised Breit-Wigner formalism did decrease the width~\cite{LHCb:2021auc}.}. The total widths from the strong and
radiative decays for the isospin singlet and triplet state read,
\begin{eqnarray}
    \text{Isospin singlet: }    \Gamma_{\text{str}}+\Gamma_\text{EM} =46.7^{+2.7}_{-2.9} \text{ keV},~\label{eq:wd_iso1} \\
    \text{Isospin triplet: }    \Gamma_{\text{str}}+\Gamma_\text{EM} =31.2^{+2.2}_{-2.4} \text{ keV}.~\label{eq:wd_iso3}
\end{eqnarray}
 The decay widths for the isospin singlet and triplet assignments are smaller than the experimental data. Therefore, one can expect that, with the improving of the measurement resolution, the decay width of $T_{cc}^+$ would be in accordance with a $D^{*+}D^0$-dominated bound state rather than the isospin triplet or the singlet. In other words, large isospin violation for $T_{cc}^+$ is supported by the present experimental results.

 In the right subfigure of Fig.~\ref{fig:4to6}, we present the ratios of  different partial decay widths. One can see that the ratio of $\Gamma[T_{cc}^+\to D^+D^0\pi^0]/\Gamma[T_{cc}^+\to D^0D^0\pi^+]$ is sensitive to the mixing angle when the angle is in the range of $(-\pi/4,\pi/4)$. When the bound state is approaching to the isospin singlet (triplet), the ratio will  increase (decrease). Meanwhile, if the $T_{cc}^+$ is dominated by the $D^{*+}D^0$ component, the radiative decay will be extremely suppressed, because the contribution from the most important diagram ($r_b$) is suppressed by the $\sin^2\theta$ in the coupling constants.

\section{Summary}~\label{sec:sum}

In this work, we study the strong and radiative decays of the newly
reported doubly charmed $T_{cc}^+$ state. The $T_{cc}^+$ state is
very close to the threshold $D^*D$. It seems to be a sibling of
$X(3872)$ in the double-charm systems.  Its long-range structure is
very important due to the quite small binding energy. In the
molecular scheme, we investigate the kinetics-allowed strong decays
$T_{cc}^+\to D^0D^0\pi^+$, $T_{cc}^+\to D^+D^0\pi^0$ and radiative
decays $D^+D^0 \gamma$, which are sensitive to the long-range
structure of $T_{cc}^+$.

In our calculations, we include the $D^{*+}D^0$ and $D^{*0}D^+$ as
two channels rather than presuming prior isospin assignment. We
adopt a well-used coupled-channel effective field theory, which is
cutoff-independent and satisfies the renormalization group
invariance. We extract the coupling constants of $T_{cc}^+$ to
$D^{*+}D^0$ and $D^{*0}D^+$  channels from the residuals of the
$T$-matrix. We relate the coupling constants to the wave functions
in the Schr\"odinger equation. Our results show the coupling
constants depend on both the binding energy and the mixing angle of
the two channels. With the coupling constants and the  strong and
radiation vertices of $D^*$ mesons from experiments, we obtain the
strong and radiative decay widths of $T_{cc}^+$. Our numerical
results show that the decay width of $T_{cc}^+\to D^0D^0\pi^+$ is
the largest one, which is consistent with the experimental
observation. We also find the theoretical total strong and radiative
width will approach the experimental value in the single channel
limit (pure $D^{*+}D^0$ component), which reads
$\Gamma_{\text{str}}+\Gamma_\text{EM} =59.7^{+4.6}_{-4.4} \text{ keV}$. Thus, we can infer that the mixing angle would be very small.
If the $T_{cc}^+$ is the pure
$D^{*+}D^0$ molecule, the radiative decay width is very tiny, which
is less likely to be detected in the near future. The ratio of
$\Gamma[T_{cc}^+\to D^+D^0\pi^0]/\Gamma[T_{cc}^+\to D^0D^0\pi^+]$ is
sensitive to the mixing angle when the angle is in the range of
$(-\pi/4,\pi/4)$. Therefore, it can be used to judge the proportion
of  $D^0D^{\ast+}$ and $D^+D^{\ast0}$ inside the $T_{cc}^+$.

Our results do not depend on the cutoff parameter. The isospin
violation effect is rigorously considered in coupled-channel
formalism, and all the relevant uncertainties are seriously
estimated. Once the new experimental results for the decays of
$T_{cc}^+$ are available, one can easily read out its inner
structure information from Figs.~\ref{fig:1to3} and \ref{fig:4to6}.
Unlike the $X(3872)$, there is no hidden-charm channel [e.g.,
$J/\psi \rho$, $J/\psi \omega$ and $\chi_{c1}(2P)$ channels for
$X(3872)$] interference to $T_{cc}^+$, so this state can also give
us a very clean platform to uncover the interaction details between
a pair of charmed mesons.

After this work, the LHCb Collaboration released the analysis within the unitarised Breit-Wigner formalism~\cite{LHCb:2021auc}. One can see their results in Eq.~\eqref{eq:lhcb_2} are in accordance with ours in Eqs.~\eqref{eq:wd_single}, \eqref{eq:wd_iso1} and \eqref{eq:wd_iso3}.

\begin{appendix}

\section{Amplitude calculation}~\label{app:amplt}
We first use Fig.~\ref{fig:feyn}($s_a$) as an example to illustrate
the calculation of the strong decay. The $D^{\ast+}\to D^+\pi^0$
amplitude is $\mathcal{A}=g_\pi q_\pi \cdot \epsilon_{D^*}$, where
$\epsilon_{D^*}$ and $q_\pi$ are the polarization vector of $D^*$
mesons and momentum of pion, respectively. The differences of
$g_\pi$ extracted from $D^{\ast+}\to D^+\pi^0$ and $D^{\ast+}\to
D^0\pi^+$ decays are very tiny (constrained by the isospin
symmetry)~\cite{ParticleDataGroup:2020ssz}. We take the averaged
value of coupling constant $g_\pi\simeq11.9$ as our input. Since
the $D^{\ast0}$ width is still unknown, we assume the isospin
symmetry and use the same coupling constant as that of the
$D^{\ast+}$. Finally, the amplitude of $T_{cc}^+\to
D^{+}D^{0}\pi^{0}$ reads
\begin{eqnarray}
{\cal A}[T_{cc}^+\to D^{+}D^{0}\pi^{0}]
=\frac{g_{1}\epsilon_{T}^{\mu}(g_{\mu\nu}-\frac{p_{12\mu}p_{12\nu}}{m_{D^{*}}^{2}})g_\pi
p_{2}^\nu}{p_{12}^{2}-m_{D^{*}}^{2}+im_{D^{*}}\Gamma_{D^{*}}},
\end{eqnarray}
where $\epsilon_{T}^{\mu}$ represents the polarization vector of
$T_{cc}^+$. $p_{12}$ and $p_2$ stand for the momenta of the $D^{*+}$
and $\pi^0$, respectively. $\Gamma_{D^{*}}$ is the width of $D^\ast$
meson.

We then use the Fig.~\ref{fig:feyn}($r_a$) to illustrate the
calculation of radiative decay amplitude. The radiative decay vertex
of $D^{*}\to D\gamma$ can be parameterized as follows,
 \begin{eqnarray}
            {\cal A}[D^{*}\to D\gamma]=g_{\gamma}\varepsilon_{\mu\nu\alpha\beta}\epsilon_{\gamma}^{\mu}p_{D^{*}}^{\nu}p_{\gamma}^{\alpha}\epsilon_{D^{*}}^{\beta},
\end{eqnarray}
where $g_\gamma$ denotes the effective coupling constant. Its value
is extracted from the partial decay widths of $D^{\ast+,0}\to
D^{+,0}\gamma$~\cite{ParticleDataGroup:2020ssz}, respectively. For
the $D^{\ast0}$ meson, we take its total width as a range
$40$$-$$80$ keV, which covers the most of the theoretical results,
e.g.~\cite{Ebert:2002xz,Choi:2007se,Becirevic:2009xp,Wang:2019mhm}.
Then the $T_{cc}^+\to D^{+}D^{0}\gamma$ amplitude reads
\begin{eqnarray}
{\cal A}[T_{cc}^+&\to& D^{+}D^{0}\gamma]\nonumber\\
&=&\frac{g_{1}\epsilon_{T}^{\mu}(g_{\mu\nu}-\frac{p_{12\mu}p_{12\nu}}{m_{D^{*}}^{2}})g_{\gamma}\varepsilon_{\rho\sigma\alpha\nu}\epsilon_{\gamma}^{\rho}p_{D^{*}}^{\sigma}p_{\gamma}^{\alpha}}{p_{12}^{2}-m_{D^{*}}^{2}+im_{D^{*}}\Gamma_{D^{*}}}.
\end{eqnarray}

\begin{figure}[!htbp]
    \centering  \includegraphics[width=0.32\textwidth]{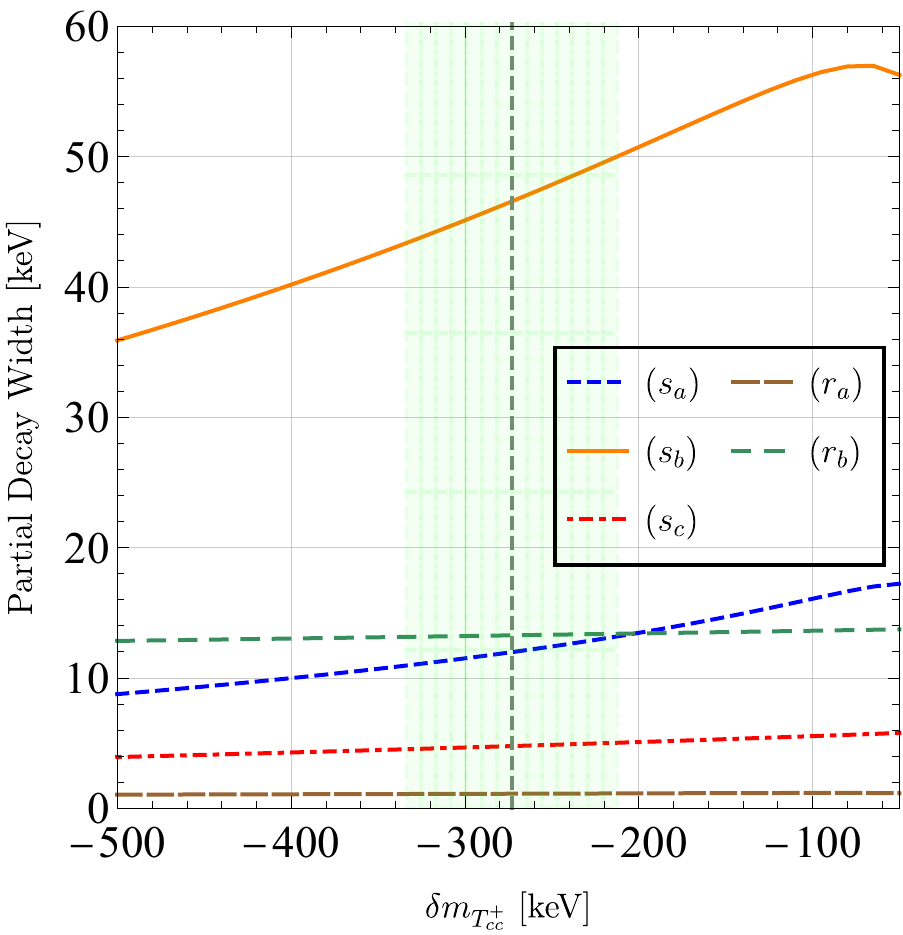}
    \caption{The contribution  to the partial widths of $T_{cc}^+$ from each Feynman diagram, where the interferences between diagrams with the same final states are switched off. The $\cos \theta$ and $\sin \theta$ in coupling constants of Eq.~\eqref{eq:g1g2} are both set to be $1$.}\label{fig:alldiag}
\end{figure}

In order to identify the dominant diagrams, we estimate the
contribution of each diagram by switching off the interference
effect and replace the $\cos \theta$ and $\sin \theta$ with $1$. The
results in Fig.~\ref{fig:alldiag} show that the ($s_b$) and ($r_b$)
are the dominant diagrams contributing to the strong and radiative
decays, respectively, which are almost $4$ times larger than the
contributions from other diagrams. The strong decay width arising
from ($s_b$) is also much larger than the radiative one from
($r_b$).

\end{appendix}


\begin{acknowledgements}
We are grateful to the helpful discussions with Prof.~Eulogio Oset
and Dr.~Rui Chen. We also thank Mikhail Mikhasenko for helpful discussions.  This project was supported by the National
Natural Science Foundation of China (11975033 and 12070131001). This
project was also funded by the Deutsche Forschungsgemeinschaft (DFG,
German Research Foundation, Project ID 196253076-TRR 110). G.J. Wang
was supported by JSPS KAKENHI (No.20F20026). B. W. was supported by
the Start-up Funds for Young Talents of Hebei University
(No. 521100221021).
\end{acknowledgements}

\bibliography{ref}

\end{document}